%% file: CavityPaperArxiv.tex
\RequirePackage{fixltx2e}

\documentclass[reprint,amsmath,amssymb,aps,]{revtex4-1}

\usepackage{graphicx}
\usepackage{dcolumn}
\usepackage{bm}
\usepackage{placeins}
\usepackage{dblfloatfix}
\usepackage{upgreek}

\begin{document}

\preprint{APS/123-QED}

\title{Cavity-enhanced frequency up-conversion in rubidium vapour}

\author{Rachel F Offer$^1$}
\author{Johnathan W C Conway$^1$}
\author{Erling Riis$^1$}
\author{Sonja Franke-Arnold$^2$}
\author{Aidan S Arnold$^1$}
\email{aidan.arnold@strath.ac.uk}
\affiliation{
 $^1$Department of Physics, SUPA, University of Strathclyde, Glasgow G4 0NG, United Kingdom\\
 $^2$School of Physics and Astronomy, SUPA, University of Glasgow, Glasgow G12 8QQ, United Kingdom
}

\date{\today}

\begin{abstract}
We report the first use of a ring cavity to both enhance the output power and dramatically narrow the linewidth (${<1\,}$MHz) of blue light generated by four wave mixing in a rubidium vapour cell.  We find that the high output power available in our cavity-free system leads to power broadening of the generated blue light linewidth.  Our ring cavity removes this limitation, allowing high output power and narrow linewidth to be achieved concurrently.  As the cavity blue light is widely tunable over the $^{85}$Rb 5S$_{1/2}$ F=3 $\rightarrow$ 6P$_{3/2}$ transition, this narrow linewidth light would be suitable for near-resonant rubidium studies including, for example, second-stage laser cooling. 
\end{abstract}

\maketitle

Atomic vapours are a versatile tool for studying a wide range of nonlinear phenomena.  In particular, quasi-resonant atomic systems allow processes such as electromagnetically induced transparency, fast and slow light, lasing without inversion and four wave mixing (FWM) to be studied at low light intensities \cite{Fleischhauer2005}.  The resonant enhancement of FWM in a rubidium vapour is such that it can be used for efficient frequency up-conversion of near-infrared light ($780\,$nm and $776\,$nm) to blue light ($420\,$nm)  \cite{Zibrov2002,Akulshin2009,Meijer2006,Vernier2010}.  For optimal pump detuning, vapour pressure and pump polarisation, modest power diode laser pumps can be used to generate $1\,$mW of coherent blue light, corresponding to a conversion efficiency of $260\,$\%/W \cite{Vernier2010}.  Such efficient FWM has applications ranging from quantum information \cite{Spiller2015,Camacho2009} to second-stage laser cooling \cite{Mckay2011,Duarte2011} and sensitive atomic imaging \cite{Sheludko2008,Yang2012}.    

Recent work on FWM in rubidium systems has show that transverse phase structure, for example orbital angular momentum (OAM), can be transferred between the pump and generated beams \cite{Walker2012, Akulshin2016}.  The ability to efficiently transfer OAM between different wavelengths may be important for future applications of structured light \cite{Franke-Arnold2008}.  Efficient FWM is not restricted to this particular system and various wavelengths can be generated by making use of different atomic states \cite{Becerra2008,Sell2014,Akulshin2014a} or different alkali metals \cite{Schultz2009}.  High conversion efficiencies have also been demonstrated in rubidium-filled hollow-core photonic crystal fibers \cite{Londero2009}.

\begin{figure}[b]
\centering
\includegraphics[width=\linewidth]{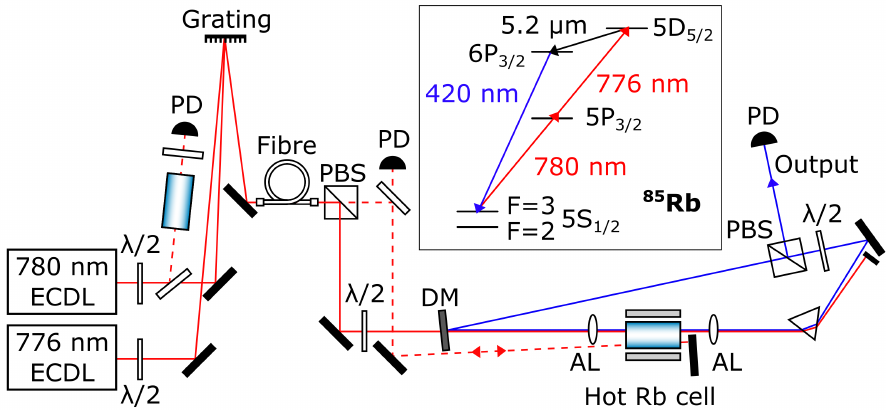}
\caption{Schematic of the experimental setup.  Abbreviations used are: PD (photodiode), PBS (polarising beam splitter), DM (dichroic mirror) and AL (achromatic lens, f $=200\,$mm).  Dashed lines represent spectroscopy probe beams used to monitor the $780\,$nm and $776\,$nm detunings.}
\label{fig:setup}
\end{figure} 


In this letter we investigate the effect of adding a ring cavity, singly resonant with the generated blue light, to our rubidium vapour FWM system \cite{Vernier2010}.  We find that a low finesse cavity more than doubles the output power and greatly reduces the linewidth of the blue light produced.  In previous single pass FWM experiments, for low output powers (around $10\,\upmu$W), the linewidth of the coherent blue light has been reported to be ${\leq3\,}$MHz \cite{Zibrov2002,Akulshin2014}.  However, in our single pass setup, up to $340\,\upmu$W of coherent emission can be generated.  For these high output powers the linewidth of the blue light increases to around $33\,$MHz.  This increase in linewidth is consistent with power broadening of the $420\,$nm transition due to the high peak blue light intensity, as discussed later in this work.  Adding a ring cavity imposes stringent spectral coherence, allowing blue light to be generated with high output power ($940\,\upmu$W) as well as a narrow linewidth (${\leq1\,}$MHz).  FWM in a ring cavity using a purely near-infrared FWM scheme within rubidium has also recently been investigated \cite{Mikhailov2015}.


Our experimental set-up and the relevant level scheme for $420\,$nm light generation in a rubidium vapour is shown in figure \ref{fig:setup}.  The $780\,$nm and $776\,$nm pump beams undergo a single pass through a heated rubidium cell, exciting the two photon resonance between the 5S$_{1/2}$ ground state and the 5D$_{5/2}$ excited state.  This develops a population inversion on the 5D$_{5/2}$ $\rightarrow$ 6P$_{3/2}$ transition which produces a $5.2\,\upmu$m field via amplified spontaneous emission (ASE) \cite{Zibrov2002}. This initial ASE together with the pump lasers establishes three photon coherence on the 5S$_{1/2}$ $\rightarrow$ 6P$_{3/2}$ transition, which in turn allows for the coherent emission of $420\,$nm light via FWM.  The ring cavity is designed to be singly resonant with this generated blue light, adding a strong constraint on the blue light frequency. The cavity also enhances the effective length of the "laser medium", thereby increasing FWM conversion efficiency. One would expect that the cavity also has an impact on the phase matching conditions for the FWM process, which we will investigate in the future.

The $780\,$nm and $776\,$nm pump beams are provided by two free running extended-cavity diode lasers (ECDLs).  To ensure they are copropagating the pump beams are overlapped on a grating and then coupled into a polarisation maintaining single mode optical fibre.  The combined $780\,$nm and $776\,$nm fibre output is then horizontally polarised before entering the cavity through a dichroic mirror.  Two achromatic lenses form a 2\textit{f} imaging system (${f=200\,}$mm) that focusses the near-IR pump beam to a $e^{-2}$ radius of $52\,\upmu$m in the centre of the heated rubidium cell.  The cell is $25\,$mm long and contains $^{87}$Rb and $^{85}$Rb in their natural abundancies.  The cell vapour temperature was determined to within $\pm1\,^\circ$C using absorption spectroscopy of a weak ($1\,\upmu$W) collimated $780\,$nm probe beam \cite{Siddons2008}.  

FWM within the rubidium vapour produces horizontally polarised $420\,$nm light, copropagating with the pump beams.  Light at $5.2\,\upmu$m is also generated \cite{Akulshin2014a} but it is not observed in our setup as it is absorbed by the glass cell.  In order for the cavity to be singly-resonant with the blue light we use a prism to separate the $420\,$nm light from the near-IR pump beam.  The pump beam is then blocked and the blue light is fed back to the heated cell.  A half waveplate and a polarising beam splitter (PBS) are used to couple light out of the cavity.  The waveplate allows the amount of output coupling to be controlled.  We have studied the effect of the cavity on FWM for $65\,\%$ and $5\,\%$ output coupling, which correspond to a cavity finesse of 3.5 and 12.8 respectively.  The parasitic loss in our cavity is around $25\,\%$, the majority of this is due to loss at the PBS and the four $4\,\%$ reflections at the cell.  

\begin{figure}[b!]
\begin{minipage}{\textwidth}
\centering
\includegraphics[width=1 \linewidth]{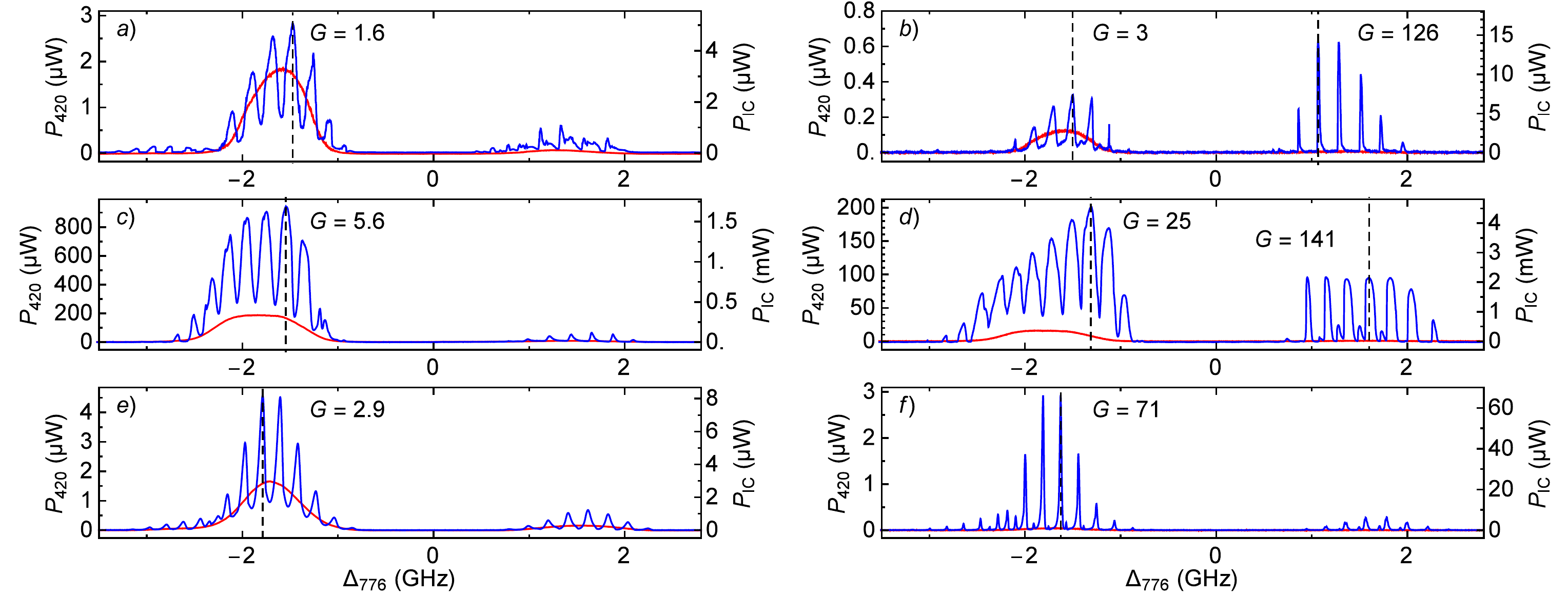}
\caption{Blue output power as a function of $776\,$nm detuning for a single pass ($P_{SP}$, red lines) and with the cavity ($P_C$, blue lines).  The right hand scale shows the blue intracavity power, $P_{IC}$.  Plots (a-f) correspond to the following conditions: (a, b) $130\,^\circ$C, $1.6\,$mW $780\,$nm, $2.7\,$mW $776\,$nm; (c, d) $130\,^\circ$C, $13\,$mW $780\,$nm, $23\,$mW $776\,$nm; (e, f) $90\,^\circ$C, $13\,$mW $780\,$nm, $23\,$mW $776\,$nm.  Cell temperatures of $130\,^\circ$C and $90\,^\circ$C correspond to vapour pressures of $0.12\,$Pa and $0.009\,$Pa respectively.  The output coupling at the PBS was $65\,\%$ for (a, c and e) and $5\,\%$ in (b, d and f).  Absolute frequency scales are accurate to $\pm0.1\,$GHz.  The $780\,$nm detuning, chosen to maximise single pass blue power, was (a, b) $1.7\,$GHz; (c, d) $1.8\,$GHz; (e, f) $1.6\,$GHz.  Representative values of the gain, $G = P_{C}/P_{SP}$, are shown, with the detuning each value was calculated at marked by a vertical dashed line.}
\label{fig:spectra}
\end{minipage}
\end{figure}

Firstly, we will discuss the effect of the cavity on blue output power.  We do this by comparing the output power as a function of $776\,$nm detuning for single pass FWM and with-cavity FWM, shown by the red and blue curves in figure \ref{fig:spectra}.  The output power was monitored using a photodiode at the cavity output and the single pass results were recorded simply by blocking the cavity after the PBS.  When recording spectra the $780\,$nm laser was set to the detuning for maximum single pass blue power, as detailed in figure \ref{fig:spectra}.  The $776\,$nm detuning was determined by $780\,$nm and $776\,$nm two photon spectroscopy in the heated Rb cell and is given relative to the $^{85}$Rb 5P$_{3/2}$ F=4 $\rightarrow$ 5D$_{5/2}$ F=5 transition.   The $780\,$nm detuning (relative to the $^{85}$Rb 5S$_{1/2}$ F=3 $\rightarrow$ 5P$_{3/2}$ F=4 transition) was determined by saturated absorption spectroscopy in a room temperature Rb cell.  

For single pass FWM, as the $776\,$nm laser is scanned across the $^{85}$Rb 5P$_{3/2}$ $\rightarrow$ 5D$_{5/2}$ transition, there are two detunings for which blue light is produced, near $\Delta_{776}=-1.8\,$GHz and $\Delta_{776}=1.2\,$GHz.  These correspond to two-photon resonance with the 5S$_{1/2}$ F = 3 $\rightarrow$ 5D$_{5/2}$ and 5S$_{1/2}$ F = 2 $\rightarrow$ 5D$_{5/2}$ transitions respectively.  In the cavity-enhanced results this same behaviour is observed but with the addition of large increases in blue output power when the $420\,$nm light, whose frequency scans with the $776\,$nm frequency, is resonant with the cavity.  

Due to energy conservation, the frequency of the FWM fields must satisfy the condition $\omega_{780}+\omega_{776} = \omega_{5200} + \omega_{420}$, where $\omega_{780}$, $\omega_{776}$, $\omega_{5200}$ and $\omega_{420}$ are the frequency of the $780\,$nm, $776\,$nm, $5200\,$nm and $420\,$nm fields respectively.  As a result, if the frequency of either of the pump lasers is changed then $\omega_{420}$ or $\omega_{5200}$ (or both) must 

\clearpage
\noindent change accordingly.  For the case of near resonant stepwise excitation of the 5S$_{1/2}$ $\rightarrow$ 5D$_{5/2}$ transition it has been shown that the $420\,$nm frequency exactly mirrors changes in pump frequency \cite{Akulshin2012}, suggesting that the $5.2\,\upmu$m field remains resonant with the atomic transition.  However, if the pump lasers are far off resonance ($1\,$THz) with the 5P$_{3/2}$ state then it has been reported that $\omega_{5200}$ can vary as well as $\omega_{420}$ \cite{Brekke2015}.  In our experiment we have measured (using a scanning Fabry-Perot interferometer) the change in $\omega_{420}$ due to a change in $\omega_{776}$ to be given by $\Delta\omega_{420}=0.92(1) \Delta\omega_{776}$.  This indicates that $\omega_{5200}$ and $\omega_{420}$ have mutual tuning consistent with relative Doppler shifts, as seen in Ref. \cite{Brekke2015}, likely due to the detuning of our pump lasers from the 5P$_{3/2}$ intermediate state (${<2\,}$GHz).  The mean separation of the observed cavity resonances in figure \ref{fig:spectra} is $198(2)\,$MHz.  This corresponds to a change in $420\,$nm frequency of $182(2)\,$MHz, which is in strong agreement with the expected free spectral range based on the length of our cavity, $181.9(3)\,$MHz.




The result shown in figure \ref{fig:spectra} (c) was obtained for the conditions we determined optimal for single pass FWM: a cell temperature of $130\,^\circ$C and maximum available pump powers ($13\,$mW $780\,$nm, $23\,$mW $776\,$nm).  In this case $340\,\upmu$W of single pass blue light is generated directly after the cell.  The cavity output coupling was set to $65\,\%$.  This value was chosen as it produced maximum cavity-enhanced output power.  Under these conditions a peak of $940\,\upmu$W of output power was produced at the cavity output; 2.8 times the power after the cell for a single pass.  

\FloatBarrier

To fully understand the with-cavity trace in figure \ref{fig:spectra} (c) a theoretical model of the system is required.  Such a model is beyond the scope of this letter, but in order to ascertain which processes would be of particular importance to such a model we have recorded spectra for a wide range of conditions: reduced cell temperature [figure \ref{fig:spectra} (e) and (f)], reduced pump power [figure \ref{fig:spectra} (a) and (b)] and decreased output coupling (right hand column of figure \ref{fig:spectra}).


\begin{figure}[b!]
\centering
\includegraphics[width= \linewidth]{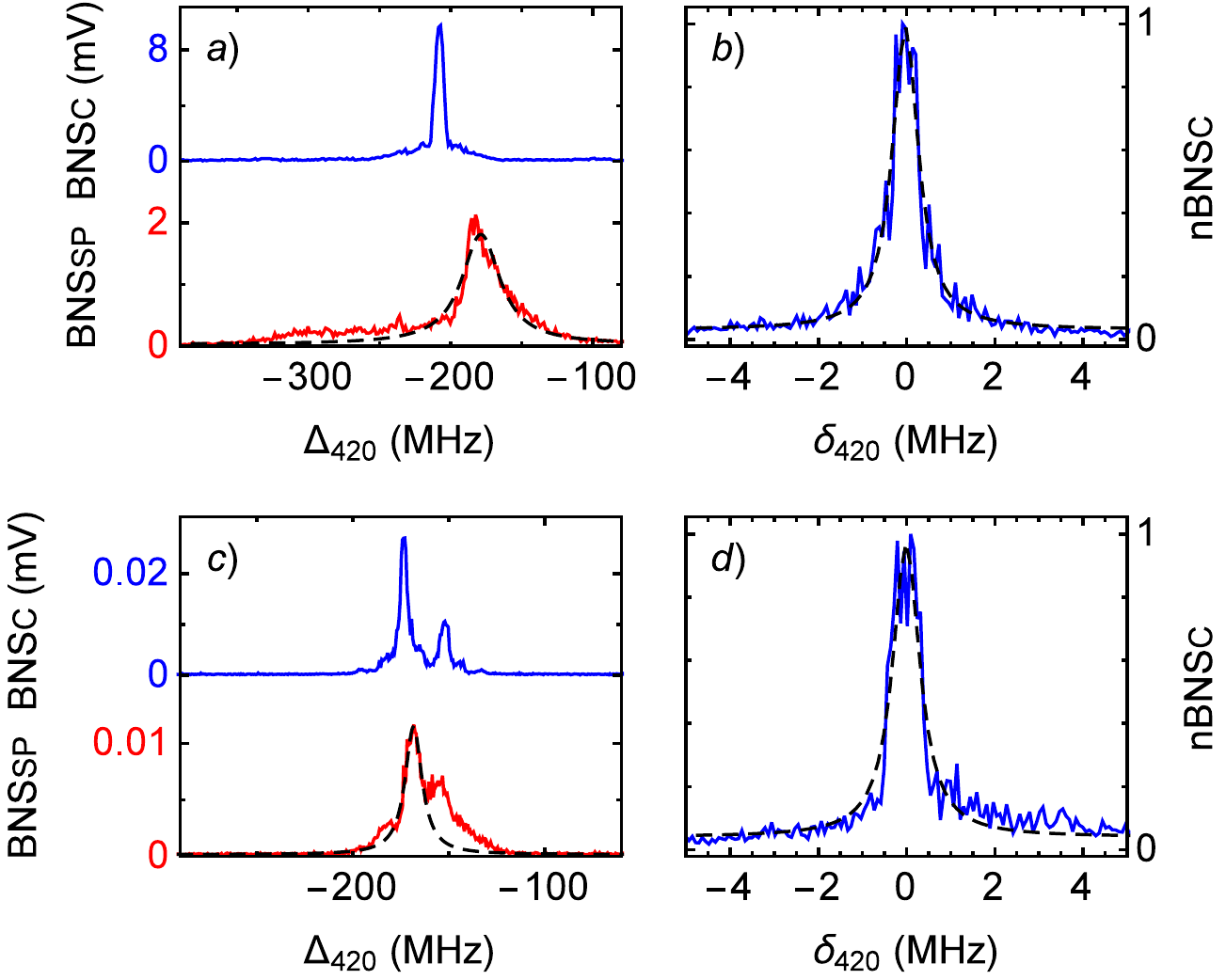}
\caption{Beat note between the FWM blue light and a $420\,$nm ECDL.  The FWM cavity conditions were: $65\,\%$ output coupling, $13\,$mW 780 nm, $23\,$mW 776 nm and cell temperature (a, b) $130\,^\circ$C and (c, d) $90\,^\circ$C.  The 780 nm and 776 nm detunings were within $0.1\,$GHz of their optimal detunings for single pass FWM.  Plots (a) and (c) compare the beat note signal (BNS) for a single-pass (red, BNSsp) and with the cavity (blue, BNSc), $\Delta_{420}$ gives the detuning of the FWM blue light from the $^{85}$Rb 5S$_{1/2}$ F=3 $\rightarrow$ 6P$_{3/2}$ F=4 transition, to within $\pm25\,$MHz.  In (a) and (c) the width of the cavity-enhanced BNS is limited by the signal analyser sweep rate; (b) and (d) show the normalised BNSc (nBNSc) over a smaller scan range on a relative frequency scale.  Image (d) shows only the larger of the two peaks in (c).  The dashed lines are Lorentzian fits with FWHM (a) $33\,$MHz; (b) $0.7\,$MHz; (c) $11\,$MHz and (d) $0.7\,$MHz.}
\label{fig:beatnote}
\end{figure} 

Firstly, we note that for high conversion efficiency, as in figure \ref{fig:spectra} (c) and (d), saturation effects become important, as can be seen by the flattening off of the resonances.  This is particularly obvious for positive detuning in figure \ref{fig:spectra} (d).  Another feature of the high conversion efficiency traces is that the cavity enhances the blue output power even when it appears to be off resonant.  As $\Delta_{776}$ changes, the $420\,$nm light will come in and out of resonance with the cavity.  One would therefore expect the with-cavity output power to vary above and below the single pass output power, unlike in figure \ref{fig:spectra} (c) and (d).  Both reducing the pump power, figure \ref{fig:spectra} (a) and (b), and reducing the cell temperature, figure \ref{fig:spectra} (e) and (f), cause the minimum cavity-enhanced power to be less than the single pass power, as expected.   

By comparing the width of the resonances in each of the traces in figure \ref{fig:spectra} the predominant broadening mechanisms can be determined.  Both reduced cell temperature and reduced pump power result in a narrowing of resonances.  This suggests that power broadening of the $420\,$nm transition dominates over collision broadening.  Comparing the left and right column in figure \ref{fig:spectra} it is also clear that the resonance width of the "passive" cavity (determined by the finesse) also plays a role.  For $65\,\%$ output coupling (left column) the passive cavity resonance width is $51\,$MHz, whilst for $5\,\%$ output coupling (right column) it is reduced to $14\,$MHz.  In figure \ref{fig:spectra} (e) and (f) this decrease is enough to outweigh the increase in power broadening due to the increased intracavity power, and so the resonances in (f) are narrower.  In figure \ref{fig:spectra} (c) and (d) however the opposite is true, the change in power broadening is largest and consequently the resonances are broader for reduced output coupling. 


However, it is clear that power broadening and the cavity finesse are not the only broadening mechanisms.  For example, figure \ref{fig:spectra} (a) has wider peaks than figure \ref{fig:spectra} (e), for similar intracavity power, suggesting that collision broadening may have some contribution.  Moreover, in figure \ref{fig:spectra} (b) the cavity resonances near the F = 2 two-photon transition (positive detuning) are narrower and give much higher gain than those near the F = 3 transition (negative detuning).  A theoretical model of the system will undoubtedly provide insight into this behaviour.


We have also demonstrated that the cavity significantly decreases the linewidth of the generated blue light.  We obtained the linewidth by beating the FWM blue beam against a $420\,$nm ECDL (Newport Vantage tunable diode laser) and measured the resulting beat note using a spectrum analyser.  The frequency of the $420\,$nm ECDL was monitored using saturated absorption spectroscopy, allowing the absolute frequency of the blue FWM light to be determined as well as the linewidth.  Figure \ref{fig:beatnote} shows the result of the beat note measurement for single pass and cavity-enhanced blue light generation, taken at both $90\,$$^\circ$C and $130\,$$^\circ$C.  The $780\,$nm and $776\,$nm pump powers were $13\,$mW and $23\,$mW respectively.  In the following we will first briefly discuss the beat note obtained for blue light generated via single pass FWM, and then go on to discuss the cavity-enhanced case.



The single pass beat note, both at $90\,^\circ$C and $130\,^\circ$C, is composed of more than one subpeak.  Similar substructure has been observed previously in \cite{Zibrov2002} where it was explained by the 6P$_{3/2}$ hyperfine splitting of $10$, $20$ and $40\,$MHz between the F'$=1,2,3,4$ levels.  However, in our measurement the width of the subpeaks makes it difficult to determine if the substructure we observe is from the same source.  Fitting to one of the subpeaks of the observed beat note signal gives the beat note linewidth to be $11\,$MHz ($33\,$MHz) at $90\,^\circ$C ($130\,^\circ$C).  We attribute this difference in FWHM to power broadening of the $420\,$nm transition, as collision broadening of the $420\,$nm transition is negligible. Based on the single pass blue power and the $e^{-2}$ radius of the blue light in the rubidium cell ($46\upmu$m), we calculate the power broadened width of the $420\,$nm transition to be $4\,$MHz ($41\,$MHz) at $90\,^\circ$C ($130\,^\circ$C) \cite{Boyd}. 


For a single pass, we find that maximal blue light is generated slightly red detuned from the $420\,$nm transition, as shown in figure \ref{fig:beatnote} (a) and (c).  These detunings are within a Doppler width of the $420\,$nm resonance and so are in agreement with previous work \cite{Akulshin2012}.  In addition, we find that the blue light can be tuned easily to frequencies either side of the transition, with a FWHM tuning range of $920(25)\,$MHz and $770(25)\,$MHz at $130\,^\circ$C and $90\,^\circ$C respectively.  

\FloatBarrier


Figure \ref{fig:beatnote} (b) and (d) show the beat note for the cavity-enhanced blue light.  At both $130\,^\circ$C and $90\,^\circ$C the linewidth is dramatically narrowed to ${\leq1\,}$MHz FWHM.  For the $130\,^\circ$C cell the beat note produced is a single sharp peak of FWHM $0.7\,$MHz.  At $90\,^\circ$C the with-cavity light has a similar linewidth but there is an additional secondary peak in the beat note signal.  As this beat note is for the cavity-enhanced case this secondary peak is likely due to a cavity mode, rather than due to hyperfine processes, indeed measurements of the beam profile at the cavity output indicate that it may be due to higher order transverse modes.   

The beat note linewidth will have some contribution from the linewidth of the $420\,$nm ECDL used for the beat note measurement.  To estimate this contribution the linewidth of the ECDL was measured separately by recording the frequency noise at the side of a Doppler broadened transmission feature.  The autocorrelation of the laser frequency noise at $0.1\,$ms was found to be $0.5\,$MHz.  The $0.1\,$ms time scale is relevant as it corresponds to the time taken for the spectrum analyser to scan over the FWHM of the with-cavity beat note signals.  This suggests that a large proportion of the beat note FWHM may come from the ECDL linewidth, and therefore the linewidth of the cavity-enhanced FWM light will be much less than $0.7\,$MHz on a $0.1\,$ms time scale.  On the same timescale, the autocorrelation of the $780\,$nm and $776\,$nm pump lasers are $0.2\,$MHz and $0.6\,$MHz respectively. This suggests that the cavity is able to narrow the linewidth of the FWM light to less than the total linewidth of the pump fields.

In conclusion, we have demonstrated the first use of a ring cavity to both enhance the power output and dramatically narrow the linewidth of blue light generated via FWM in a rubidium vapour cell.  For a cell temperature of $130\,^\circ$C the resulting output power is nearly $1\,$mW (nearly three times the output power of the cavity-free case) and the linewidth drops from a power broadened $33\,$MHz to less than $1\,$MHz.  Furthermore, the blue light is generated with a frequency close to the $^{85}$Rb 5S$_{1/2}$ $\rightarrow$ 6P$_{3/2}$ transition and is tuneable over a FWHM range of almost $1\,$GHz.  The increased output power, narrow linewidth and large tuning range could make this FWM in a ring cavity system a valuable light source for efficient $^{85}$Rb Bose-Einstein condensate production \cite{Cornish2000}.  In addition, if the input laser powers were increased or the large parasitic losses present in our cavity minimised, for example by using an anti-reflection coated or Brewster cell, then even larger output powers would be possible.  The datasets used in this Letter are available via Ref. \cite{data}.   

\bigskip

\section*{Funding Information}

Leverhulme Trust (RPG-2013-386); InnovateUK QuDOS.

\bigskip
\noindent

\input{CavityPaperArxiv.bbl}

\end{document}

%% file: CavityPaperArxiv.bbl
%

%% file: CavityPaperArxiv.bbl
\begin{thebibliography}{27}%
\makeatletter
\providecommand \@ifxundefined [1]{%
 \@ifx{#1\undefined}
}%
\providecommand \@ifnum [1]{%
 \ifnum #1\expandafter \@firstoftwo
 \else \expandafter \@secondoftwo
 \fi
}%
\providecommand \@ifx [1]{%
 \ifx #1\expandafter \@firstoftwo
 \else \expandafter \@secondoftwo
 \fi
}%
\providecommand \natexlab [1]{#1}%
\providecommand \enquote  [1]{``#1''}%
\providecommand \bibnamefont  [1]{#1}%
\providecommand \bibfnamefont [1]{#1}%
\providecommand \citenamefont [1]{#1}%
\providecommand \href@noop [0]{\@secondoftwo}%
\providecommand \href [0]{\begingroup \@sanitize@url \@href}%
\providecommand \@href[1]{\@@startlink{#1}\@@href}%
\providecommand \@@href[1]{\endgroup#1\@@endlink}%
\providecommand \@sanitize@url [0]{\catcode `\\12\catcode `\$12\catcode
  `\&12\catcode `\#12\catcode `\^12\catcode `\_12\catcode `\%12\relax}%
\providecommand \@@startlink[1]{}%
\providecommand \@@endlink[0]{}%
\providecommand \url  [0]{\begingroup\@sanitize@url \@url }%
\providecommand \@url [1]{\endgroup\@href {#1}{\urlprefix }}%
\providecommand \urlprefix  [0]{URL }%
\providecommand \Eprint [0]{\href }%
\providecommand \doibase [0]{http://dx.doi.org/}%
\providecommand \selectlanguage [0]{\@gobble}%
\providecommand \bibinfo  [0]{\@secondoftwo}%
\providecommand \bibfield  [0]{\@secondoftwo}%
\providecommand \translation [1]{[#1]}%
\providecommand \BibitemOpen [0]{}%
\providecommand \bibitemStop [0]{}%
\providecommand \bibitemNoStop [0]{.\EOS\space}%
\providecommand \EOS [0]{\spacefactor3000\relax}%
\providecommand \BibitemShut  [1]{\csname bibitem#1\endcsname}%
\let\auto@bib@innerbib\@empty
\bibitem [{\citenamefont {Fleischhauer}\ \emph {et~al.}(2005)\citenamefont
  {Fleischhauer}, \citenamefont {Imamoglu},\ and\ \citenamefont
  {Marangos}}]{Fleischhauer2005}%
  \BibitemOpen
  \bibfield  {author} {\bibinfo {author} {\bibfnamefont {M.}~\bibnamefont
  {Fleischhauer}}, \bibinfo {author} {\bibfnamefont {A.}~\bibnamefont
  {Imamoglu}}, \ and\ \bibinfo {author} {\bibfnamefont {J.~P.}\ \bibnamefont
  {Marangos}},\ }\href@noop {} {\bibfield  {journal} {\bibinfo  {journal} {Rev.
  Mod. Phys.}\ }\textbf {\bibinfo {volume} {77}},\ \bibinfo {pages} {633}
  (\bibinfo {year} {2005})}\BibitemShut {NoStop}%
\bibitem [{\citenamefont {Zibrov}\ \emph {et~al.}(2002)\citenamefont {Zibrov},
  \citenamefont {Lukin}, \citenamefont {Hollberg},\ and\ \citenamefont
  {Scully}}]{Zibrov2002}%
  \BibitemOpen
  \bibfield  {author} {\bibinfo {author} {\bibfnamefont {A.~S.}\ \bibnamefont
  {Zibrov}}, \bibinfo {author} {\bibfnamefont {M.~D.}\ \bibnamefont {Lukin}},
  \bibinfo {author} {\bibfnamefont {L.}~\bibnamefont {Hollberg}}, \ and\
  \bibinfo {author} {\bibfnamefont {M.~O.}\ \bibnamefont {Scully}},\ }\href
  {\doibase 10.1103/PhysRevA.65.051801} {\bibfield  {journal} {\bibinfo
  {journal} {Phys. Rev. A}\ }\textbf {\bibinfo {volume} {65}},\ \bibinfo
  {pages} {051801} (\bibinfo {year} {2002})}\BibitemShut {NoStop}%
\bibitem [{\citenamefont {Akulshin}\ \emph {et~al.}(2009)\citenamefont
  {Akulshin}, \citenamefont {McLean}, \citenamefont {Sidorov},\ and\
  \citenamefont {Hannaford}}]{Akulshin2009}%
  \BibitemOpen
  \bibfield  {author} {\bibinfo {author} {\bibfnamefont {A.~M.}\ \bibnamefont
  {Akulshin}}, \bibinfo {author} {\bibfnamefont {R.~J.}\ \bibnamefont
  {McLean}}, \bibinfo {author} {\bibfnamefont {A.~I.}\ \bibnamefont {Sidorov}},
  \ and\ \bibinfo {author} {\bibfnamefont {P.}~\bibnamefont {Hannaford}},\
  }\href {\doibase 10.1364/OE.17.022861} {\bibfield  {journal} {\bibinfo
  {journal} {Opt. Express}\ }\textbf {\bibinfo {volume} {17}},\ \bibinfo
  {pages} {22861} (\bibinfo {year} {2009})},\ \Eprint
  {http://arxiv.org/abs/0910.2292} {arXiv:0910.2292} \BibitemShut {NoStop}%
\bibitem [{\citenamefont {Meijer}\ \emph {et~al.}(2006)\citenamefont {Meijer},
  \citenamefont {White}, \citenamefont {Smeets}, \citenamefont {Jeppesen},\
  and\ \citenamefont {Scholten}}]{Meijer2006}%
  \BibitemOpen
  \bibfield  {author} {\bibinfo {author} {\bibfnamefont {T.}~\bibnamefont
  {Meijer}}, \bibinfo {author} {\bibfnamefont {J.~D.}\ \bibnamefont {White}},
  \bibinfo {author} {\bibfnamefont {B.}~\bibnamefont {Smeets}}, \bibinfo
  {author} {\bibfnamefont {M.}~\bibnamefont {Jeppesen}}, \ and\ \bibinfo
  {author} {\bibfnamefont {R.~E.}\ \bibnamefont {Scholten}},\ }\href {\doibase
  10.1364/OL.31.001002} {\bibfield  {journal} {\bibinfo  {journal} {Opt.
  Lett.}\ }\textbf {\bibinfo {volume} {31}},\ \bibinfo {pages} {1002} (\bibinfo
  {year} {2006})}\BibitemShut {NoStop}%
\bibitem [{\citenamefont {Vernier}\ \emph {et~al.}(2010)\citenamefont
  {Vernier}, \citenamefont {Franke-Arnold}, \citenamefont {Riis},\ and\
  \citenamefont {Arnold}}]{Vernier2010}%
  \BibitemOpen
  \bibfield  {author} {\bibinfo {author} {\bibfnamefont {A.}~\bibnamefont
  {Vernier}}, \bibinfo {author} {\bibfnamefont {S.}~\bibnamefont
  {Franke-Arnold}}, \bibinfo {author} {\bibfnamefont {E.}~\bibnamefont {Riis}},
  \ and\ \bibinfo {author} {\bibfnamefont {A.~S.}\ \bibnamefont {Arnold}},\
  }\href {\doibase 10.1364/OE.18.017020} {\bibfield  {journal} {\bibinfo
  {journal} {Opt. Express}\ }\textbf {\bibinfo {volume} {18}},\ \bibinfo
  {pages} {17020} (\bibinfo {year} {2010})},\ \Eprint
  {http://arxiv.org/abs/0911.0812} {arXiv:0911.0812} \BibitemShut {NoStop}%
\bibitem [{\citenamefont {Spiller}\ \emph {et~al.}(2005)\citenamefont
  {Spiller}, \citenamefont {Munro}, \citenamefont {Barrett},\ and\
  \citenamefont {Kok}}]{Spiller2015}%
  \BibitemOpen
  \bibfield  {author} {\bibinfo {author} {\bibfnamefont {T.~P.}\ \bibnamefont
  {Spiller}}, \bibinfo {author} {\bibfnamefont {W.~J.}\ \bibnamefont {Munro}},
  \bibinfo {author} {\bibfnamefont {S.~D.}\ \bibnamefont {Barrett}}, \ and\
  \bibinfo {author} {\bibfnamefont {P.}~\bibnamefont {Kok}},\ }\href {\doibase
  10.1080/00107510500293261} {\bibfield  {journal} {\bibinfo  {journal}
  {Contemp. Phys.}\ }\textbf {\bibinfo {volume} {46}},\ \bibinfo {pages} {407}
  (\bibinfo {year} {2005})}\BibitemShut {NoStop}%
\bibitem [{\citenamefont {Camacho}\ \emph {et~al.}(2009)\citenamefont
  {Camacho}, \citenamefont {Vudyasetu},\ and\ \citenamefont
  {Howell}}]{Camacho2009}%
  \BibitemOpen
  \bibfield  {author} {\bibinfo {author} {\bibfnamefont {R.~M.}\ \bibnamefont
  {Camacho}}, \bibinfo {author} {\bibfnamefont {P.~K.}\ \bibnamefont
  {Vudyasetu}}, \ and\ \bibinfo {author} {\bibfnamefont {J.~C.}\ \bibnamefont
  {Howell}},\ }\href {\doibase 10.1038/NPHOTON.2008.290} {\bibfield  {journal}
  {\bibinfo  {journal} {Nat. Photonics}\ }\textbf {\bibinfo {volume} {3}},\
  \bibinfo {pages} {103} (\bibinfo {year} {2009})}\BibitemShut {NoStop}%
\bibitem [{\citenamefont {McKay}\ \emph {et~al.}(2011)\citenamefont {McKay},
  \citenamefont {Jervis}, \citenamefont {Fine}, \citenamefont {Simpson-Porco},
  \citenamefont {Edge},\ and\ \citenamefont {Thywissen}}]{Mckay2011}%
  \BibitemOpen
  \bibfield  {author} {\bibinfo {author} {\bibfnamefont {D.~C.}\ \bibnamefont
  {McKay}}, \bibinfo {author} {\bibfnamefont {D.}~\bibnamefont {Jervis}},
  \bibinfo {author} {\bibfnamefont {D.~J.}\ \bibnamefont {Fine}}, \bibinfo
  {author} {\bibfnamefont {J.~W.}\ \bibnamefont {Simpson-Porco}}, \bibinfo
  {author} {\bibfnamefont {G.~J.~A.}\ \bibnamefont {Edge}}, \ and\ \bibinfo
  {author} {\bibfnamefont {J.~H.}\ \bibnamefont {Thywissen}},\ }\href@noop {}
  {\bibfield  {journal} {\bibinfo  {journal} {Phys. Rev. A}\ }\textbf {\bibinfo
  {volume} {84}},\ \bibinfo {pages} {063420} (\bibinfo {year}
  {2011})}\BibitemShut {NoStop}%
\bibitem [{\citenamefont {Duarte}\ \emph {et~al.}(2011)\citenamefont {Duarte},
  \citenamefont {Hart}, \citenamefont {Hitchcock}, \citenamefont {Corcovilos},
  \citenamefont {Yang}, \citenamefont {Reed},\ and\ \citenamefont
  {Hulet}}]{Duarte2011}%
  \BibitemOpen
  \bibfield  {author} {\bibinfo {author} {\bibfnamefont {P.~M.}\ \bibnamefont
  {Duarte}}, \bibinfo {author} {\bibfnamefont {R.~A.}\ \bibnamefont {Hart}},
  \bibinfo {author} {\bibfnamefont {J.~M.}\ \bibnamefont {Hitchcock}}, \bibinfo
  {author} {\bibfnamefont {T.~A.}\ \bibnamefont {Corcovilos}}, \bibinfo
  {author} {\bibfnamefont {T.-L.}\ \bibnamefont {Yang}}, \bibinfo {author}
  {\bibfnamefont {A.}~\bibnamefont {Reed}}, \ and\ \bibinfo {author}
  {\bibfnamefont {R.~G.}\ \bibnamefont {Hulet}},\ }\href@noop {} {\bibfield
  {journal} {\bibinfo  {journal} {Phys. Rev. A}\ }\textbf {\bibinfo {volume}
  {84}},\ \bibinfo {pages} {061406(R)} (\bibinfo {year} {2011})}\BibitemShut
  {NoStop}%
\bibitem [{\citenamefont {Sheludko}\ \emph {et~al.}(2008)\citenamefont
  {Sheludko}, \citenamefont {Bell}, \citenamefont {Anderson}, \citenamefont
  {Hofmann}, \citenamefont {Vredenbregt},\ and\ \citenamefont
  {Scholten}}]{Sheludko2008}%
  \BibitemOpen
  \bibfield  {author} {\bibinfo {author} {\bibfnamefont {D.~V.}\ \bibnamefont
  {Sheludko}}, \bibinfo {author} {\bibfnamefont {S.~C.}\ \bibnamefont {Bell}},
  \bibinfo {author} {\bibfnamefont {R.}~\bibnamefont {Anderson}}, \bibinfo
  {author} {\bibfnamefont {C.~S.}\ \bibnamefont {Hofmann}}, \bibinfo {author}
  {\bibfnamefont {E.~J.~D.}\ \bibnamefont {Vredenbregt}}, \ and\ \bibinfo
  {author} {\bibfnamefont {R.~E.}\ \bibnamefont {Scholten}},\ }\href {\doibase
  10.1103/PhysRevA.77.033401} {\bibfield  {journal} {\bibinfo  {journal} {Phys.
  Rev. A}\ }\textbf {\bibinfo {volume} {77}},\ \bibinfo {pages} {033401}
  (\bibinfo {year} {2008})}\BibitemShut {NoStop}%
\bibitem [{\citenamefont {Yang}\ \emph {et~al.}(2012)\citenamefont {Yang},
  \citenamefont {Liang}, \citenamefont {He},\ and\ \citenamefont
  {Wang}}]{Yang2012}%
  \BibitemOpen
  \bibfield  {author} {\bibinfo {author} {\bibfnamefont {B.}~\bibnamefont
  {Yang}}, \bibinfo {author} {\bibfnamefont {Q.}~\bibnamefont {Liang}},
  \bibinfo {author} {\bibfnamefont {J.}~\bibnamefont {He}}, \ and\ \bibinfo
  {author} {\bibfnamefont {J.}~\bibnamefont {Wang}},\ }\href@noop {} {\bibfield
   {journal} {\bibinfo  {journal} {Opt. Express}\ }\textbf {\bibinfo {volume}
  {20}},\ \bibinfo {pages} {11944} (\bibinfo {year} {2012})}\BibitemShut
  {NoStop}%
\bibitem [{\citenamefont {Walker}\ \emph {et~al.}(2012)\citenamefont {Walker},
  \citenamefont {Arnold},\ and\ \citenamefont {Franke-Arnold}}]{Walker2012}%
  \BibitemOpen
  \bibfield  {author} {\bibinfo {author} {\bibfnamefont {G.}~\bibnamefont
  {Walker}}, \bibinfo {author} {\bibfnamefont {A.~S.}\ \bibnamefont {Arnold}},
  \ and\ \bibinfo {author} {\bibfnamefont {S.}~\bibnamefont {Franke-Arnold}},\
  }\href {\doibase 10.1103/PhysRevLett.108.243601} {\bibfield  {journal}
  {\bibinfo  {journal} {Phys. Rev. Lett.}\ }\textbf {\bibinfo {volume} {108}},\
  \bibinfo {pages} {243601} (\bibinfo {year} {2012})},\ \Eprint
  {http://arxiv.org/abs/1203.1520} {arXiv:1203.1520} \BibitemShut {NoStop}%
\bibitem [{\citenamefont {Akulshin}\ \emph {et~al.}(2016)\citenamefont
  {Akulshin}, \citenamefont {Novikova}, \citenamefont {Mikhailov},
  \citenamefont {Suslov},\ and\ \citenamefont {McLean}}]{Akulshin2016}%
  \BibitemOpen
  \bibfield  {author} {\bibinfo {author} {\bibfnamefont {A.~M.}\ \bibnamefont
  {Akulshin}}, \bibinfo {author} {\bibfnamefont {I.}~\bibnamefont {Novikova}},
  \bibinfo {author} {\bibfnamefont {E.~E.}\ \bibnamefont {Mikhailov}}, \bibinfo
  {author} {\bibfnamefont {S.~A.}\ \bibnamefont {Suslov}}, \ and\ \bibinfo
  {author} {\bibfnamefont {R.~J.}\ \bibnamefont {McLean}},\ }\href@noop {}
  {\bibfield  {journal} {\bibinfo  {journal} {Opt. Lett.}\ }\textbf {\bibinfo
  {volume} {41}},\ \bibinfo {pages} {1146} (\bibinfo {year}
  {2016})}\BibitemShut {NoStop}%
\bibitem [{\citenamefont {Franke-Arnold}\ \emph {et~al.}(2008)\citenamefont
  {Franke-Arnold}, \citenamefont {Allen},\ and\ \citenamefont
  {Padgett}}]{Franke-Arnold2008}%
  \BibitemOpen
  \bibfield  {author} {\bibinfo {author} {\bibfnamefont {S.}~\bibnamefont
  {Franke-Arnold}}, \bibinfo {author} {\bibfnamefont {L.}~\bibnamefont
  {Allen}}, \ and\ \bibinfo {author} {\bibfnamefont {M.}~\bibnamefont
  {Padgett}},\ }\href {\doibase 10.1002/lpor.200810007} {\bibfield  {journal}
  {\bibinfo  {journal} {Laser Photon. Rev.}\ }\textbf {\bibinfo {volume} {2}},\
  \bibinfo {pages} {299} (\bibinfo {year} {2008})}\BibitemShut {NoStop}%
\bibitem [{\citenamefont {Becerra}\ \emph {et~al.}(2008)\citenamefont
  {Becerra}, \citenamefont {Willis}, \citenamefont {Rolston},\ and\
  \citenamefont {Orozco}}]{Becerra2008}%
  \BibitemOpen
  \bibfield  {author} {\bibinfo {author} {\bibfnamefont {F.~E.}\ \bibnamefont
  {Becerra}}, \bibinfo {author} {\bibfnamefont {R.~T.}\ \bibnamefont {Willis}},
  \bibinfo {author} {\bibfnamefont {S.~L.}\ \bibnamefont {Rolston}}, \ and\
  \bibinfo {author} {\bibfnamefont {L.~A.}\ \bibnamefont {Orozco}},\ }\href
  {\doibase 10.1103/PhysRevA.78.013834} {\bibfield  {journal} {\bibinfo
  {journal} {Phys. Rev. A}\ }\textbf {\bibinfo {volume} {78}},\ \bibinfo
  {pages} {013834} (\bibinfo {year} {2008})}\BibitemShut {NoStop}%
\bibitem [{\citenamefont {Sell}\ \emph {et~al.}(2014)\citenamefont {Sell},
  \citenamefont {Gearba}, \citenamefont {DePaola},\ and\ \citenamefont
  {Knize}}]{Sell2014}%
  \BibitemOpen
  \bibfield  {author} {\bibinfo {author} {\bibfnamefont {J.~F.}\ \bibnamefont
  {Sell}}, \bibinfo {author} {\bibfnamefont {M.~A.}\ \bibnamefont {Gearba}},
  \bibinfo {author} {\bibfnamefont {B.~D.}\ \bibnamefont {DePaola}}, \ and\
  \bibinfo {author} {\bibfnamefont {R.~J.}\ \bibnamefont {Knize}},\ }\href
  {\doibase 10.1364/OL.39.000528} {\bibfield  {journal} {\bibinfo  {journal}
  {Opt. Lett.}\ }\textbf {\bibinfo {volume} {39}},\ \bibinfo {pages} {528}
  (\bibinfo {year} {2014})},\ \Eprint {http://arxiv.org/abs/arXiv:1311.4626v1}
  {arXiv:arXiv:1311.4626v1} \BibitemShut {NoStop}%
\bibitem [{\citenamefont {Akulshin}\ \emph
  {et~al.}(2014{\natexlab{a}})\citenamefont {Akulshin}, \citenamefont
  {Budker},\ and\ \citenamefont {McLean}}]{Akulshin2014a}%
  \BibitemOpen
  \bibfield  {author} {\bibinfo {author} {\bibfnamefont {A.}~\bibnamefont
  {Akulshin}}, \bibinfo {author} {\bibfnamefont {D.}~\bibnamefont {Budker}}, \
  and\ \bibinfo {author} {\bibfnamefont {R.}~\bibnamefont {McLean}},\ }\href
  {\doibase 10.1364/OL.39.000845} {\bibfield  {journal} {\bibinfo  {journal}
  {Opt. Lett.}\ }\textbf {\bibinfo {volume} {39}},\ \bibinfo {pages} {845}
  (\bibinfo {year} {2014}{\natexlab{a}})},\ \Eprint
  {http://arxiv.org/abs/1311.0071} {arXiv:1311.0071} \BibitemShut {NoStop}%
\bibitem [{\citenamefont {Schultz}\ \emph {et~al.}(2009)\citenamefont
  {Schultz}, \citenamefont {Abend}, \citenamefont {D\"{o}ring}, \citenamefont
  {Debs}, \citenamefont {Altin}, \citenamefont {White}, \citenamefont
  {Robins},\ and\ \citenamefont {Close}}]{Schultz2009}%
  \BibitemOpen
  \bibfield  {author} {\bibinfo {author} {\bibfnamefont {J.~T.}\ \bibnamefont
  {Schultz}}, \bibinfo {author} {\bibfnamefont {S.}~\bibnamefont {Abend}},
  \bibinfo {author} {\bibfnamefont {D.}~\bibnamefont {D\"{o}ring}}, \bibinfo
  {author} {\bibfnamefont {J.~E.}\ \bibnamefont {Debs}}, \bibinfo {author}
  {\bibfnamefont {P.~A.}\ \bibnamefont {Altin}}, \bibinfo {author}
  {\bibfnamefont {J.~D.}\ \bibnamefont {White}}, \bibinfo {author}
  {\bibfnamefont {N.~P.}\ \bibnamefont {Robins}}, \ and\ \bibinfo {author}
  {\bibfnamefont {J.~D.}\ \bibnamefont {Close}},\ }\href {\doibase
  10.1364/OL.34.002321} {\bibfield  {journal} {\bibinfo  {journal} {Opt.
  Lett.}\ }\textbf {\bibinfo {volume} {34}},\ \bibinfo {pages} {2321} (\bibinfo
  {year} {2009})},\ \Eprint {http://arxiv.org/abs/0905.3980} {arXiv:0905.3980}
  \BibitemShut {NoStop}%
\bibitem [{\citenamefont {Londero}\ \emph {et~al.}(2009)\citenamefont
  {Londero}, \citenamefont {Venkataraman}, \citenamefont {Bhagwat},
  \citenamefont {Slepkov},\ and\ \citenamefont {Gaeta}}]{Londero2009}%
  \BibitemOpen
  \bibfield  {author} {\bibinfo {author} {\bibfnamefont {P.}~\bibnamefont
  {Londero}}, \bibinfo {author} {\bibfnamefont {V.}~\bibnamefont
  {Venkataraman}}, \bibinfo {author} {\bibfnamefont {A.~R.}\ \bibnamefont
  {Bhagwat}}, \bibinfo {author} {\bibfnamefont {A.~D.}\ \bibnamefont
  {Slepkov}}, \ and\ \bibinfo {author} {\bibfnamefont {A.~L.}\ \bibnamefont
  {Gaeta}},\ }\href {\doibase 10.1103/PhysRevLett.103.043602} {\bibfield
  {journal} {\bibinfo  {journal} {Phys. Rev. Lett.}\ }\textbf {\bibinfo
  {volume} {103}},\ \bibinfo {pages} {043602} (\bibinfo {year}
  {2009})}\BibitemShut {NoStop}%
\bibitem [{\citenamefont {Akulshin}\ \emph
  {et~al.}(2014{\natexlab{b}})\citenamefont {Akulshin}, \citenamefont
  {Perrella}, \citenamefont {Truong}, \citenamefont {Luiten}, \citenamefont
  {Budker},\ and\ \citenamefont {McLean}}]{Akulshin2014}%
  \BibitemOpen
  \bibfield  {author} {\bibinfo {author} {\bibfnamefont {A.}~\bibnamefont
  {Akulshin}}, \bibinfo {author} {\bibfnamefont {C.}~\bibnamefont {Perrella}},
  \bibinfo {author} {\bibfnamefont {G.-W.}\ \bibnamefont {Truong}}, \bibinfo
  {author} {\bibfnamefont {A.}~\bibnamefont {Luiten}}, \bibinfo {author}
  {\bibfnamefont {D.}~\bibnamefont {Budker}}, \ and\ \bibinfo {author}
  {\bibfnamefont {R.}~\bibnamefont {McLean}},\ }\href {\doibase
  10.1007/s00340-014-5823-0} {\bibfield  {journal} {\bibinfo  {journal} {Appl.
  Phys. B}\ }\textbf {\bibinfo {volume} {117}},\ \bibinfo {pages} {203}
  (\bibinfo {year} {2014}{\natexlab{b}})}\BibitemShut {NoStop}%
\bibitem [{\citenamefont {Mikhailov}\ \emph {et~al.}(2014)\citenamefont
  {Mikhailov}, \citenamefont {Evans}, \citenamefont {Budker}, \citenamefont
  {Rochester},\ and\ \citenamefont {Novikova}}]{Mikhailov2015}%
  \BibitemOpen
  \bibfield  {author} {\bibinfo {author} {\bibfnamefont {E.~E.}\ \bibnamefont
  {Mikhailov}}, \bibinfo {author} {\bibfnamefont {J.}~\bibnamefont {Evans}},
  \bibinfo {author} {\bibfnamefont {D.}~\bibnamefont {Budker}}, \bibinfo
  {author} {\bibfnamefont {S.~M.}\ \bibnamefont {Rochester}}, \ and\ \bibinfo
  {author} {\bibfnamefont {I.}~\bibnamefont {Novikova}},\ }\href@noop {}
  {\bibfield  {journal} {\bibinfo  {journal} {Opt. Eng.}\ }\textbf {\bibinfo
  {volume} {53}},\ \bibinfo {pages} {102709} (\bibinfo {year}
  {2014})}\BibitemShut {NoStop}%
\bibitem [{\citenamefont {Siddons}\ \emph {et~al.}(2008)\citenamefont
  {Siddons}, \citenamefont {Adams}, \citenamefont {Ge},\ and\ \citenamefont
  {Hughes}}]{Siddons2008}%
  \BibitemOpen
  \bibfield  {author} {\bibinfo {author} {\bibfnamefont {P.}~\bibnamefont
  {Siddons}}, \bibinfo {author} {\bibfnamefont {C.~S.}\ \bibnamefont {Adams}},
  \bibinfo {author} {\bibfnamefont {C.}~\bibnamefont {Ge}}, \ and\ \bibinfo
  {author} {\bibfnamefont {I.~G.}\ \bibnamefont {Hughes}},\ }\href {\doibase
  10.1088/0953-4075/41/15/155004} {\bibfield  {journal} {\bibinfo  {journal}
  {J. Phys. B: At. Mol. Opt. Phys.}\ }\textbf {\bibinfo {volume} {41}},\
  \bibinfo {pages} {155004} (\bibinfo {year} {2008})}\BibitemShut {NoStop}%
\bibitem [{\citenamefont {Akulshin}\ \emph {et~al.}(2012)\citenamefont
  {Akulshin}, \citenamefont {Perrella}, \citenamefont {Truong}, \citenamefont
  {McLean},\ and\ \citenamefont {Luiten}}]{Akulshin2012}%
  \BibitemOpen
  \bibfield  {author} {\bibinfo {author} {\bibfnamefont {A.}~\bibnamefont
  {Akulshin}}, \bibinfo {author} {\bibfnamefont {C.}~\bibnamefont {Perrella}},
  \bibinfo {author} {\bibfnamefont {G.-W.}\ \bibnamefont {Truong}}, \bibinfo
  {author} {\bibfnamefont {R.}~\bibnamefont {McLean}}, \ and\ \bibinfo {author}
  {\bibfnamefont {A.}~\bibnamefont {Luiten}},\ }\href {\doibase
  10.1088/0953-4075/45/24/245503} {\bibfield  {journal} {\bibinfo  {journal}
  {J. Phys. B: At. Mol. Opt. Phys.}\ }\textbf {\bibinfo {volume} {45}},\
  \bibinfo {pages} {245503} (\bibinfo {year} {2012})}\BibitemShut {NoStop}%
\bibitem [{\citenamefont {Brekke}\ and\ \citenamefont
  {Herman}(2015)}]{Brekke2015}%
  \BibitemOpen
  \bibfield  {author} {\bibinfo {author} {\bibfnamefont {E.}~\bibnamefont
  {Brekke}}\ and\ \bibinfo {author} {\bibfnamefont {E.}~\bibnamefont
  {Herman}},\ }\href@noop {} {\bibfield  {journal} {\bibinfo  {journal} {Opt.
  Lett.}\ }\textbf {\bibinfo {volume} {40}},\ \bibinfo {pages} {5674} (\bibinfo
  {year} {2015})}\BibitemShut {NoStop}%
\bibitem [{\citenamefont {Boyd}(2008)}]{Boyd}%
  \BibitemOpen
  \bibfield  {author} {\bibinfo {author} {\bibfnamefont {R.~W.}\ \bibnamefont
  {Boyd}},\ }\href@noop {} {\emph {\bibinfo {title} {Nonlinear Optics}}}\
  (\bibinfo  {publisher} {Academic Press},\ \bibinfo {year} {2008})\BibitemShut
  {NoStop}%
\bibitem [{\citenamefont {Cornish}\ \emph {et~al.}(2000)\citenamefont
  {Cornish}, \citenamefont {Claussen}, \citenamefont {Roberts}, \citenamefont
  {Cornell},\ and\ \citenamefont {Wieman}}]{Cornish2000}%
  \BibitemOpen
  \bibfield  {author} {\bibinfo {author} {\bibfnamefont {S.~L.}\ \bibnamefont
  {Cornish}}, \bibinfo {author} {\bibfnamefont {N.~R.}\ \bibnamefont
  {Claussen}}, \bibinfo {author} {\bibfnamefont {J.~L.}\ \bibnamefont
  {Roberts}}, \bibinfo {author} {\bibfnamefont {E.~A.}\ \bibnamefont
  {Cornell}}, \ and\ \bibinfo {author} {\bibfnamefont {C.~E.}\ \bibnamefont
  {Wieman}},\ }\href@noop {} {\bibfield  {journal} {\bibinfo  {journal} {Phys.
  Rev. Lett.}\ }\textbf {\bibinfo {volume} {85}},\ \bibinfo {pages} {1795}
  (\bibinfo {year} {2000})}\BibitemShut {NoStop}%
\bibitem [{dat()}]{data}%
  \BibitemOpen
  \href@noop {} {}\bibinfo {howpublished}
  {http://dx.doi.org/10.15129/422007e8-5f7a-4d5b-a56f-ef458c43dcd8}\BibitemShut
  {NoStop}%
\end{thebibliography}%
